\documentclass[a4paper]{jpconf}
\usepackage{graphicx}
\usepackage{amsfonts}
\usepackage{amsmath}
\usepackage{amssymb}
\usepackage{dsfont}
\usepackage{multirow}
\usepackage{float}

\renewcommand{\Tr}{\mathop{\mathrm{Tr}}\nolimits}
\renewcommand{\Re}{\mathop{\mathrm{Re}}\nolimits}

\begin{document}
\hfill {\bf HU-EP-15/46} \\

\vspace*{-2cm}
\title{Towards the quark--gluon plasma Equation of State with 
dynamical strange and charm quarks
\footnote[99]{Contribution to the International Conference 
{\it Strangeness in Quark Matter 2015}, Dubna, July 6 -- 11, 2015, 
presented by A. Trunin}
}

\author{Florian~Burger$^{1}$, Ernst--Michael Ilgenfritz$^{2}$, Maria Paola Lombardo$^3$, \\ 
        Michael M\"uller--Preussker$^1$ and Anton Trunin$^{2}$}

\address{$^1$ Physics Department, Humboldt University Berlin, Newtonstr. 15, 12489 Berlin, Germany}
\address{$^2$ Bogoliubov Laboratory of Theoretical Physics, Joint Institute for Nuclear Research, \\ 
          ~~Joliot--Curie Str. 6, 141980 Dubna, Russia}
\address{$^3$ Frascati National Laboratory, National Institute for Nuclear Physics,\\
		~~Via Enrico Fermi 40, 00044 Frascati (Rome), Italy}

\ead{amtrnn@gmail.com}

\begin{abstract}
We present an ongoing project aimed at determining the thermodynamic Equation of State
(EoS) of quark--gluon matter from lattice QCD with two generations of dynamical quarks. 
We employ the Wilson twisted mass implementation for the fermionic fields and 
the improved Iwasaki gauge action. Relying on $T=0$ data obtained by the ETM Collaboration
the strange and charm quark masses are fixed at their physical values, while the pion
mass takes four values in the range from 470 MeV down to 210 MeV. The temperature is
varied within a fixed--lattice scale approach. The values for the pseudocritical temperature 
are obtained from various observables. For the EoS we show preliminary results for
the pure gluonic contribution obtained at the pion mass value 370 MeV, where we can
compare with previously obtained results with $N_f=2$ degenerate light flavours.
\end{abstract}

The Equation of State (EoS) plays a key role in the description of the quark--gluon
plasma, serving as important and necessary input for hydrodynamical models describing the 
space--time evolution of hot QCD matter in relativistic heavy ion collisions. At the moment, 
lattice QCD is the only known way to non--perturbatively calculate the EoS from first
principles and without application of phenomenological models, although this calculation 
is currently possible only in the limit of an (almost) vanishing baryon chemical 
potential~\cite{phil}.
Detailed EoS results were presented recently by the Budapest--Wuppertal~\cite{bw} and 
HotQCD~\cite{hotqcd} collaborations, both with $N_f=2+1$ flavours of staggered fermions. 
This type of fermions on the lattice is attractive due to the relatively low computational
costs, but is known to have certain conceptual problems from the theoretical point of view
(''rooting trick`` problem). As an alternative, one can consider a Wilson--type fermionic
action, which is theoretically safe but computationally demanding. In~this talk we report on 
the current status of the ''twisted mass at finite temperature`` (tmfT) project~\cite{burger13}
devoted in particular to the calculation of the EoS with $N_f=2+1+1$ flavours of 
twisted mass Wilson fermions. To our knowledge, this project represents a first study of the 
Equation of State with almost realistic masses of strange and charm quarks implemented with a
Wilson--type fermion discretization. In the works~\cite{bw2} and~\cite{milc}, 
where the $c$ quark was also taken into account, the staggered approach was used, 
and in Ref.~\cite{borsanyi2010} charm effects were estimated on the basis of a $N_f=2+1$ 
analysis. The recent progress for the EoS with Wilson fermions includes $N_f=2$ twisted mass 
results~\cite{burger15} and $N_f=2+1$ studies with various action improvements~\cite
{umeda,borsanyi2012}.

The Wilson twisted mass fermionic action for light and heavy quark doublets has the 
following form~\cite{tm}:
\begin{eqnarray}
\label{action}
S^\text{light}_f[U,\chi_l,{\overline\chi}_l]  &=&  \sum_{x,y} \overline{\chi}_l(x)
        \left[\delta_{x,y} -\kappa D_\mathrm{W}(x,y)[U] +
        2 i \kappa a {{\mu_l}} \gamma_5 \delta_{x,y}  \tau^3 \right ] \chi_l(y) \,,\\
\nonumber
S^\text{heavy}_f[U,\chi_h, \overline{\chi}_h]  &=&  \sum_{x,y} \overline{\chi}_h(x)
         \left[ \delta_{x,y} - \kappa D_W(x,y)[U] +
         2 i \kappa {a {\mu_{\sigma}}} \gamma_5 \delta_{x,y} \tau^1 +
         2 \kappa {a {\mu_{\delta}}} \delta_{x,y} \tau^3 \right] \chi_h(y),
\end{eqnarray}
where $D_\mathrm{W}[U]$ is the usual Wilson operator, $a$ is the lattice spacing, 
and $\chi_{l,h}$ are quark spinors in the twisted basis. The hopping parameter $\kappa$ 
is set to its coupling dependent critical value $\kappa_c(\beta)$ leading to the 
so-called ''maximal twist`` of the action~\eqref{action} with the property of 
automatic $\mathcal O(a)$ improvement for expectation values of any 
operator~\cite{frezzotti}. The parameter $\mu_l$ describes the mass of the 
degenerate light quark doublet, which is still unphysically large in our study: 
the charged pion mass values $m_{\pi^\pm}$ considered at present are 
210, 260, 370 and 470~MeV. The heavy twisted mass parameters $\mu_{\sigma}$ and 
$\mu_{\delta}$ have been tuned in the unitary approach to reproduce approximately the 
physical $K$ and $D$ meson mass values within the accuracy of 10\%, thus allowing for a 
realistic treatment of $s$ and $c$ quarks. Three lattice spacing values $a$ are 
available for the bare gauge coupling parameter $\beta=6/g^2=1.90$, 1.95 and $2.10$ 
corresponding to $a=0.094$, 0.082 and $0.065$~fm, respectively \cite{a_data}. 

The temperature $T=1/(a N_\tau)$ is varied by changing $N_\tau=3,4,\ldots,24$, the 
latter being the lattice extent in the fourth Euclidean direction ({\it fixed--scale 
approach}~\cite{umeda}). The linear spatial extent varies between $N_\sigma = 24$ and $48$, 
such that the bound of the aspect ratio $N_\sigma/N_\tau \ge 2$ is guaranteed as the worst
case. In comparison with the more often used fixed--$N_\tau$ 
approach, such a choice allows to reduce the amount of required $T=0$ computations, 
necessary to carry out the obligate subtractions in computing the EoS. 
Fortunately, we can rely on the mass parameters and on the scale setting by the 
European Twisted Mass (ETM) Collaboration~(see \cite{a_data} and references therein).

\begin{figure}[t]
\begin{center}
\includegraphics[scale=0.85]{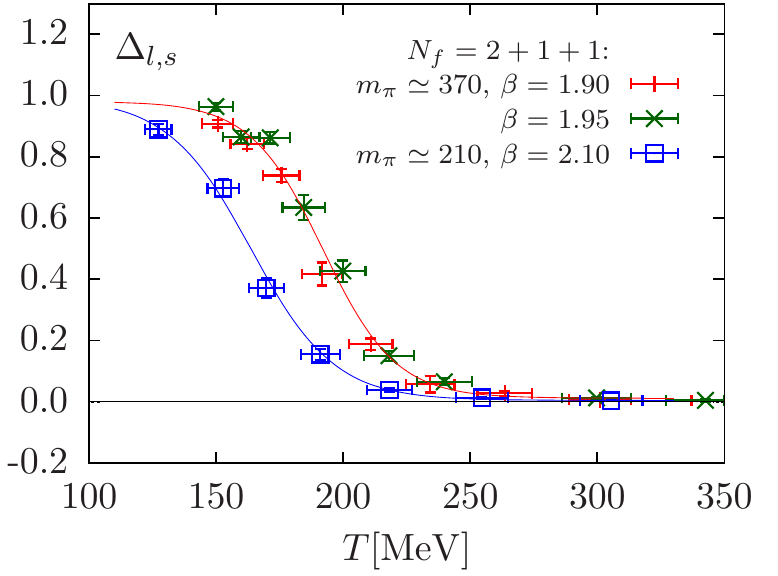} \qquad 
\includegraphics[scale=0.855]{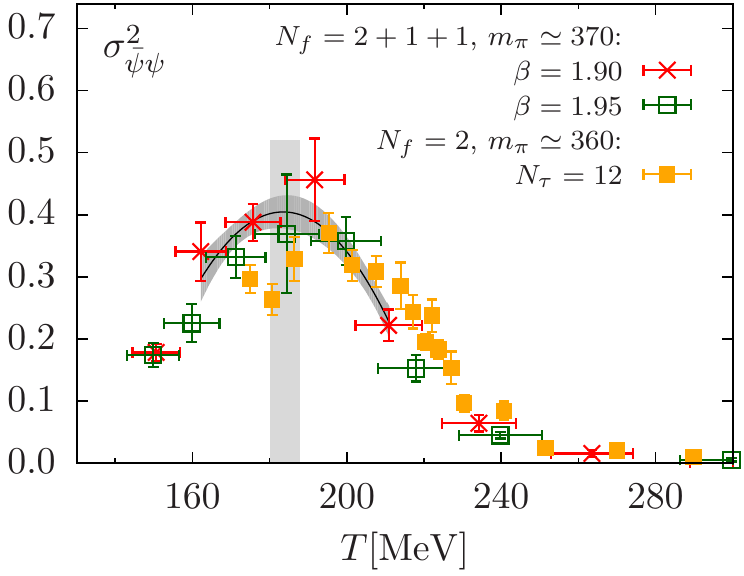} 
\caption{
$\Delta_{l,s}$ ({left}) and $\sigma^2_{\bar\psi\psi}$ ({right}) versus $T$.
}
\label{fig1}
\end{center}
\end{figure}

For determining the pseudocritical temperature(s) of the crossover region we have
computed the renormalized Polyakov loop
\begin{equation}
\langle\Re\,\mathrm L\rangle_R=\langle\Re\,\mathrm L\rangle\,\mathrm{exp}[V(r_0)/2T]\,,
\end{equation}
where $V(r_0)$ is static quark potential at the Sommer scale $r_0 \simeq 0.5~\mathrm{fm}$
and the renormalized subtracted chiral condensate
\begin{equation}
\Delta_{l,s} = \dfrac{{\langle \bar\psi\psi \rangle}_l - 
\frac{\mu_l}{\mu_s}{\langle \bar\psi\psi \rangle}_s}{{\langle \bar\psi\psi \rangle}^{T=0}_l - 
\frac{\mu_l}{\mu_s}{\langle \bar\psi\psi \rangle}^{T=0}_s}\,,
\end{equation}
where $\langle \bar\psi\psi \rangle_l$ and $\langle \bar\psi\psi \rangle_s$ are the 
light and strange quark condensates, respectively. The latter has been obtained in the 
Osterwalder--Seiler setup \cite{osterwalder, frezzotti}, which avoids 
mixing in the heavy quark sector. The mass $\mu_s$ has been determined as to reproduce the 
physical $\bar s \gamma_\mu s$--mass. Moreover, we have determined the 
unrenormalized disconnected chiral susceptibility
\begin{equation}
\sigma^2_{\bar\psi\psi}=\frac{V}{T}\bigl(\langle(\bar\psi\psi)\,^2\rangle_l
 -\langle\bar\psi\psi\rangle_l^2\bigr),
\qquad
V=a^4N^3_\sigma N_\tau.
\end{equation}
The inflexion points of  $\langle\Re\,\mathrm L\rangle_R$ and $\Delta_{l,s}$ as
well as the maxima of $\sigma^2_{\bar\psi\psi}$ versus $T$ determine the crossover
temperatures $T_L, T_\Delta$ and $T_\chi$, respectively, as shown in Table \ref{tbl}.
In Figure \ref{fig1}, as an example, we show the behaviour of $\Delta_{l,s}$ 
for the pion mass values $m_{\pi^\pm}=370$ and $210$ MeV as well as of 
$\sigma^2_{\bar\psi\psi}$ for $m_{\pi^\pm}=370$ MeV (the latter in comparison with the case
$N_f=2$ \cite{burger15}).
\begin{table}[t]
\begin{center}
\caption{\label{tbl}
Crossover temperatures for different pion mass values  $m_{\pi^\pm}$.
$N_f=2$ results are taken from Ref.~\cite{burger15} 
(all numbers in units of MeV).
}
\begin{tabular}{lllll}
\br
Number of & \multirow{2}*{$m_{\pi^\pm}$} & \multirow{2}*{$T_\chi$} & 
                     \multirow{2}*{$T_\Delta$} & \multirow{2}*{$T_{L}$}\\
flavours &&&&\\
\mr
\multirow{4}*{$N_f=2+1+1$}
 &210 & 152(5) & 164(3) & $-$\\
 &260& 170(5) & $-$     & $-$\\
 &370& 184(4) & 192(2) & 201(3)\\
 &470& 199(6) & $-$ & $-$ \\
\mr
\multirow{2}*{$N_f=2$}
 &360& 193(13) & $-$ & 219(3)(14)\\
 &430& 208(14) & $-$ & 225(3)(14)\\
\br
\end{tabular}
\end{center}
\end{table}

\begin{figure}[t]
\begin{center}
\hspace*{-1pc}
\includegraphics[scale=0.9]{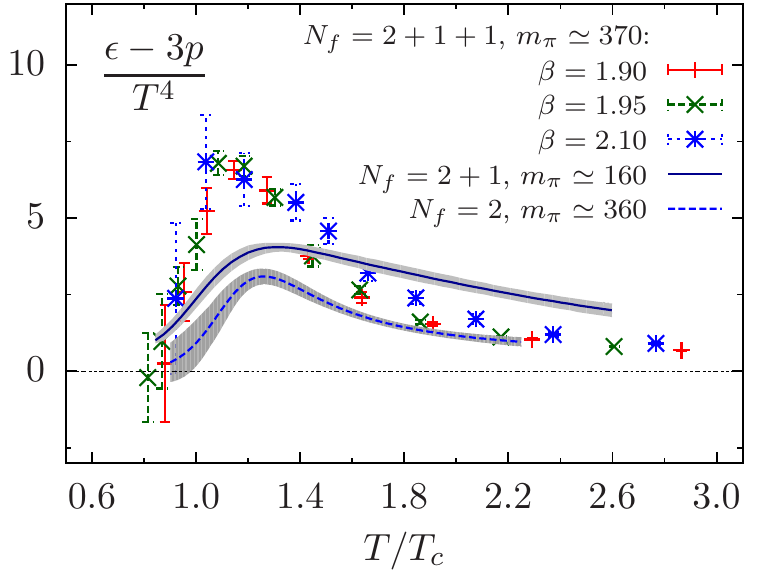}
\end{center}
\vspace*{-1pc}
\caption{\label{fig2}
\negmedspace\,Tree--level corrected gauge field contribution to the trace anomaly 
for $m_{\pi^\pm} \simeq 370$~MeV in comparison with the fitted full 
$N_f=2$ result~\cite{burger15} as well as with the continuum extrapolated
$N_f=2+1$ case~\cite{hotqcd}.}
\end{figure}

The trace anomaly or interaction measure, from which the pressure $p$ as well as
the energy density $\epsilon$ can be finally extracted as a function of $T$, is 
given by the logarithmic derivative of the lattice partition function $\mathcal Z$
\begin{equation} \label{traceanomaly}
I(T)=\epsilon-3p,\qquad \frac{I(T)}{T^4}
   =-\frac{1}{T^3V}\Bigl(\frac{d\ln\mathcal Z}{d\ln a}\Bigr)_{\rm sub}\,,
\end{equation}
where the subtracted expectation values are defined as
$\langle \ldots \rangle_{\rm sub} \equiv \langle 
\ldots \rangle_{T} - \langle \ldots \rangle_{T=0}$.
This leads to the expression
\begin{gather}
\frac{I(T)}{T^4} 
       =\frac{1}{T^3V}\sum_i\frac{db_i}{da}\bigl\langle\frac{\partial S}
                   {\partial b_i}\bigr\rangle_{\rm sub}
       = N_\tau^4\Bigl\{\bigl(- a\frac{d\beta}{da}\bigr) 
         \Bigl( \frac{c_0}{3} \langle \sum_P \Re \Tr U_P \rangle_{\rm sub} + 
                 \frac{c_1}{3}\langle \sum_R \Re \Tr U_R \rangle_{\rm sub}  \notag
       \\[-4pt]
        +\, \frac{\partial \kappa_c}{\partial \beta} 
         \langle \bar{\chi} D_\mathrm{W}[U] \chi \rangle_{\rm sub} 
         - 2a\mu_l\frac{\partial \kappa_c}{\partial \beta} 
          \langle \bar{\chi} i \gamma_5 \tau^3 \chi \rangle_{\rm sub} \Bigr) 
         + 2 \kappa_c \Bigl(a \frac{d (a\mu_l)}{da}\Bigr) \langle\bar{\chi} i \gamma_5 \tau^3 \chi \rangle_{\rm sub} 
           + \mathrm{heavy~terms}\Bigr\}\,, \notag \\ 
 I(T) = I_\text{gauge}+I_\text{light}+I_\text{heavy}
\end{gather}
with the renormalization functions ${db_i}/{da}$ to be 
fixed from ETM Collaboration $T=0$ data. Preliminary results 
for $I/T^4$ are presented in Figure \ref{fig2}, where the pure gauge field (tree--level
corrected) contribution as the most dominant one is shown for the case 
$m_{\pi^\pm} = 370$ MeV and the three lattice spacings. We compare with the full 
$N_f=2$ result at approximately the same pion mass \cite{burger15}
as well as with the fitted results for $N_f=2+1$ dynamical staggered fermion 
flavour degrees of freedom close to the physical point \cite{hotqcd}. 
The comparison demonstrates that our $N_f=2+1+1$ results turned out 
to be in the right ball park.

The computation of the light and heavy quark contributions --- being
most interesting at higher temperatures --- is under way. Their magnitude 
turns out to be rather sensitive with respect to the mass renormalization 
functions to be determined from the ETM Collaboration analysis of the 
hadronic spectrum at $T=0$.

\ack
We acknowledge generous support by the HLRN supercomputer centers Berlin and Hannover, 
the Supercomputing Center of Lomonosov Moscow State University,
as well as by the HybriLIT group of JINR.  
The work was partly financed by the Heisenberg--Landau program of BLTP JINR and 
BMBF Germany and by the ''Dynasty`` foundation.

\end{document}